\documentclass[reprint, aip, apl]{revtex4-1}
\newcount\Comments  
\usepackage{graphicx}
\usepackage{dcolumn}
\usepackage{bm}
\usepackage{xcolor} 
\usepackage[utf8]{inputenc}
\usepackage[T1]{fontenc}
\usepackage{mathptmx}
\usepackage{etoolbox}
\newcommand{\kibitz}[2]{\ifnum\Comments=1\textcolor{#1}{#2}\fi}
\usepackage{changes}
\definechangesauthor[name={Author 1}, color=red]{author1}

\makeatletter
\def\@email#1#2{%
 \endgroup
 \patchcmd{\titleblock@produce}
  {\frontmatter@RRAPformat}
  {\frontmatter@RRAPformat{\produce@RRAP{*#1\href{mailto:#2}{#2}}}\frontmatter@RRAPformat}
  {}{}
}%
\makeatother
\begin{document}

\preprint{AIP/123-QED}

\title{Sub-cycle Nanotip Field Emission of Electrons Driven by Air Plasma Generated THz Pulses}
\author{Benjamin Colmey}

\affiliation{ 
Department of Physics, McGill University, Montreal, Quebec H3A2T8, Canada
}%

\affiliation{%
Centre for the Physics of Materials, McGill University, Montreal, Canada
}%
\author{Rodrigo T. Paulino}
\affiliation{ 
Department of Physics, McGill University, Montreal, Quebec H3A2T8, Canada
}%

\affiliation{%
Centre for the Physics of Materials, McGill University, Montreal, Canada
}%

\author{{Gaspard Beaufort}}
\affiliation{ 
Department of Physics, Lund University, P.O. Box 118, 22100 Lund, Sweden
}%

\author{David G. Cooke \textsuperscript{*}}
\email{cooke@physics.mcgill.ca}

\affiliation{ 
Department of Physics, McGill University, Montreal, Quebec H3A2T8, Canada
}%

\affiliation{%
Centre for the Physics of Materials, McGill University, Montreal, Canada
}%

\date{\today}

\begin{abstract}
Terahertz pulses generated by two-color laser plasmas have reported peak field strengths exceeding MV/cm, and when illuminating metal nanotips the near-field enhancement at the tip apex should result in extremely high bunch charges and electron energies via sub-cycle cold field emission. Here, electron emission from tungsten nanotips driven by THz pulses generated by a long filament air-plasma are reported. Electron energies up to 1.1 keV and bunch charges up to 2x$10^5$ electrons per pulse were detected, well below values expected for peak field calculated via the time averaged Poynting vector. Investigations revealed a failure in the use of the time-averaged Poynting vector when applied to long filament THz pulses, due to spatio-temporal restructuring of the THz pulse in the focus. Accounting for this restructuring significantly reduces the field strength to approximately 160 ~kV/cm, consistent with the observed electron bunch charges, peak energies and their dependence on the tip position in the THz focus. Despite these findings, our results surpass previous THz plasma-driven electron generation by an order of magnitude in both electron energy and bunch charge and a path to increasing these by an additional order of magnitude by modification of the THz optics is proposed.

\end{abstract}

\maketitle

The development of sources of high brightness femtosecond electron pulses have led to significant advances in ultrafast electron microscopy and diffraction, enabling time-resolved atomic-scale imaging of dynamic processes in materials science, chemistry, and biology. \cite{Filippetto2022,Arbouet2018}  The emission of electrons from metallic surfaces and structures is often achieved by single or multi-photon photoemission and sub-cycle field emission using near-infrared femtosecond laser pulses.\cite{Ropers2007, Schenk2010, BormannPhysRevLett2010, Echternkamp2016} Early demonstrations of nanotip electron emission employed low pulse energy, high-repetition-rate near-infrared (NIR) pulses. \cite{HommelhoffPhysRevLett2006_1, HommelhoffPhysRevLett2006_2} Recently, single-cycle pulses of THz frequency light with peak fields exceeding 100~kV/cm have been used for cold-field emission (CFE) directly from the Fermi surface of metal nanotips. \cite{LiJones2016, Matte2021} The metal nanotips induce a lightning rod effect, resulting in a broadband field enhancement at the tip apex of the order $\gamma\sim\lambda/R$ where $\lambda$ is the free space wavelength and $R$ is the tip radius. \cite{houard2020nanotip, Behr2008, Podenok2006}

Sub-cycle CFE offers several advantages compared to over-the-barrier photoemission or Schottky emission, including higher peak currents, as it is not limited by material damage thresholds from photon-induced heating inherent in NIR photoemission.\cite{Wellershoff1999} Additionally, THz field emission directly accelerates electrons post-tunneling in the local field, reaching multiple keV in energy, unlike photoemission which requires acceleration through a high voltage biased region of many cm. During the flight time, the pulse broadens significantly in space and time due to space-charge effects. 
\\
\indent Pioneering work by Herink \textit{et al.} used air plasma generated THz pulses centered at 1 THz with peak fields of $\sim10$~kV/cm, \cite{herink2014} which, when aided by a -40~V bias, produced electron energies of 120~eV and yielding $\sim50$ electrons per bunch at 1~kHz from a 10~nm radius nanotip. Subsequently Li and Jones, using a 15~Hz, 20~mJ Ti:sapphire fs laser generating 450~kV/cm peak field THz pulses with a central frequency of 0.15~THz via tilted pulse front optical rectification in LiNbO$_3$, achieved electron energies up to  5~keV.\cite{LiJones2016} Most recently Matte \textit{et al.} achieved bunch charges of $10^6$ electrons per pulse at 1~kHz repetition rate with peak THz fields of 298~kV/cm and a central frequency of 1.2~THz generated by tilted pulse front optical rectification in LiNbO$_3$, reaching maximum energies of 3.5~keV.\cite{Matte2021} This electron count surpassed previous CFE results by 3 orders of magnitude and was found to be consistent with Fowler-Nordheim (FN) theory, verifying the extreme quasi-static nature of the light-matter interaction. Despite these impressive results, air-plasma sources, with reported field strengths in the MV/cm range,\cite{stepanov2003,YehAPL2007, hiroriAPL2011b, Guiramand22} hold the potential to increase both the energy and bunch charge by another order of magnitude. In this work, the capabilities of two-color air-plasma THz sources to emit high-energy electron bunches via sub-cycle field emission from metal nanotips are studied.

\begin{figure*}[ht]
\centering
\includegraphics[width=\textwidth]{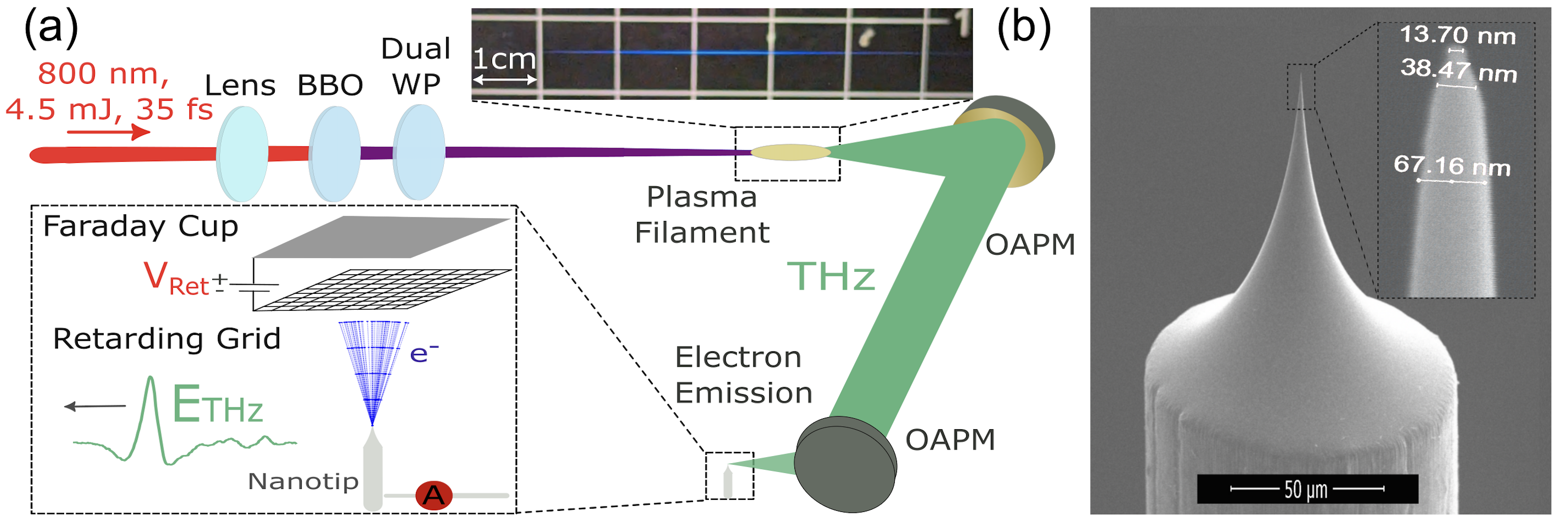} 
\caption{\label{fig:wide} (a) Diagram of the experimental setup, showing THz generation and electron emission/detection. The NIR beam is focused through a BBO crystal for 2nd harmonic generation and aligned with the 2nd harmonic by a half-wave plate, forming a plasma filament and generating THz radiation, which is then directed to the tungsten nanotip for electron emission. (b) Scanning electron microscopy image of tungsten nanotip with inset showing the apex with a tip diameter of 13.7~nm.} 

\end{figure*}

Sub-cycle tunneling of the electron in the extreme quasi-static interaction limit occurs at a rate given by the Fowler-Nordheim (FN) equation:\cite{FowlerRoySocLondon1928}

\begin{equation}
J = \frac{aF_{loc} (t)^2}{F^2_\phi \phi}e^{-{vb} \frac{F_\phi}{F_{loc}(t)}}
\end{equation}
where $a$ and $b$ are FN constants, $1.54 \times 10^6 \, \text{A eV/V}^2$ and $6.83 \times 10^9 \, \text{V eV}^{3/2}/\text{m}$ respectively,\cite{Lange2020} $F_{loc}$ is the local electric field, and $F_\phi$ is the field required to narrow the potential barrier to zero, approximately 14~GV/m for tungsten.\cite{Forbes2008} Here, $v \approx 1 - F_{loc}/ F_{\phi} $, accounts for image charges and exponentially suppresses the emission for $F_{loc} < F_\phi$.\cite{ForbesJVacSciTechB2010}
\\
\indent To reach the extreme quasi-static limit for sub-cycle CFE, the local field enhancement provided by metallic nanotips typically requires apex radii below 100~nm. These nanotips benefit from significant field enhancement at the apex due to the lightning rod effect, where the applied electric field ($F_0$) is enhanced by redistributing electric charges to cancel the field inside the conductor.\cite{Thomas2015} This effect narrows the emission area to the tip's apex, offering a distinct advantage over photoemission sources, which have emission areas as wide as the focused beam on the photocathode. This point-like emission area increases the electron beam's brightness and coherence, critical for high-resolution electron imaging.\cite{LatychevskaiaUltramicro2017} 

In this study, the experimental apparatus, shown in Fig. \ref{fig:wide} (a), made use of a Coherent Legend Elite Duo HE+ Ti:sapphire regenerative amplifier, which emits  5~mJ, 35~fs pulses at 1~kHz with a central wavelength of 795~nm. The NIR output was split with 90\% driving THz generation and 10\% to the probe line. A two-color laser plasma was generated by focusing the main beam with a 50~cm lens into a 100~$\mu$m-thick Beta-Barium-Borate (BBO) crystal placed before the focus, facilitating second harmonic generation at 400~nm. A zero-order half-wave plate after the BBO aligned the fundamental and second harmonic beams to vertically polarized states, creating a plasma filament and emitting THz radiation.

The THz beam was collimated using a 4-inch off-axis parabolic mirror (OAPM) and focused by a 3-inch OAPM onto the tungsten nanotip. Detection techniques included direct beam visualization with an IRXCAM INO384 microbolometer array THz camera, pyroelectric power measurements using a Gentec-EO THZ5B-VANTA meter, and standard electro-optic (EO) sampling for waveform and frequency spectrum analysis using a $200~\mu$m GaP crystal which was bonded to a $4$~mm inactive GaP substrate.\cite{wu1997} 

Electron emission experiments were conducted using a tungsten nanotip housed in a spherical cube vacuum chamber, maintained at $\approx 10^{-6}$~Torr. The chamber featured optical access ports and electrical feed-throughs, with the nanotip mounted on a 3-axis translation stage for precise positioning. \cite{Matte2021} Tungsten nanotips were reliably produced with tip apex radii between 10-20~nm using electrochemical etching.\cite{Lucier2004} A scanning electron microscope (SEM) image of nanotips used here is shown in Fig. \ref{fig:wide} (b). Significant reshaping of the tips was observed during previous experiments,\cite{Matte2021} leading to an estimated radius during emission between 15-35~nm. This reshaping was suggested by observed emission instability immediately after installation of a fresh tip, with the electron emission being reduced by a factor of 2 over 24 hours. At pressures of $\approx 10^{-6}$ Torr, this emission instability is likely due to the presence of N$_2$ and the formation of tungsten nitride layer on the nanotip. High fields are known to cause field-assisted etching, leading to tip reshaping.\cite{Rezeq2006}
\\
\indent A Kimball Physics model FC-73A Faraday cup detector with a  5~mm diameter aperture was placed 2~cm above the nanotip. The collection angle is then approximately 7 degrees, allowing us to detect only a subset of electrons emitted within this angle. The setup included suppression and retarding grids to filter secondary electrons and analyze the energy spectrum of the emitted electrons. Additionally, a direct electrical connection at the nanotip base allowed measurement of the total tunneling current using a Keithley 6517B electrometer. The Faraday cup permitted mesaurement of the energy spectrum of the electron beam by capturing only electrons with energy surpassing the applied retarding potential. By varying the bias voltage and observing changes in the detected current, an electron energy spectrum can be obtained. Applying a low bias voltage of -20~V to the suppression grid further reduced the detection of secondary and scattered electrons.

\begin{figure}[ht]
    \centering
    {\includegraphics[width=0.99\linewidth]{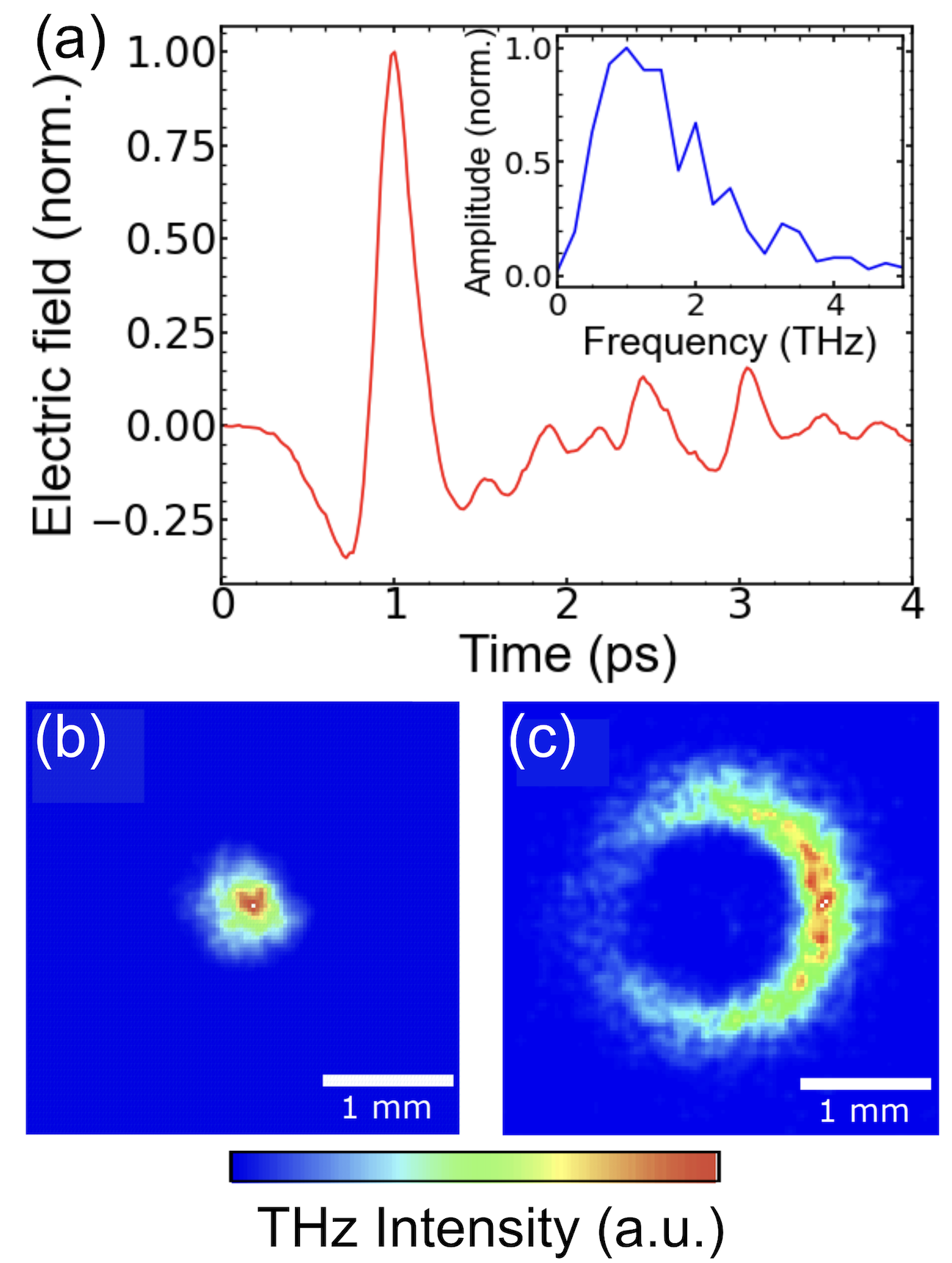}}
    \caption{(a) THz pulse waveform and Fourier spectra (inset) for THz generated from a two-color air-plasma measured via electro-optic sampling in a GaP crystal. (b)(c) THz beam profiles measured with the INO camera at 0, and 16~mm from the focus, respectively. Intensity normalized relative to each image's brightest value.}
    \label{fig:Cam}
\end{figure}

The THz emission efficiency and spatial focusing was optimized using a a pyroelectric power meter (Gentec-EO THZ5B-VANTA) and a microbolometer camera located at the THz focus. The time-domain profile of the detected THz pulse is shown in Fig. \ref{fig:Cam}(a), with the Fourier spectra in the inset showing a spectrum peaked at 1.46~THz. The FWHM of the focused beam profile, extracted from Fig. \ref{fig:Cam}(b), are $\Gamma_x = 380 \pm 120$$\mu$m and $\Gamma_y = 345 \pm 110$$\mu$m along the x- and y-axes, respectively, where the uncertainties were estimated using frequency-dependent beam waist variation and camera absorption profiles. As shown in Fig. \ref{fig:Cam}(c), at a distance of 16 mm after the focus, the beam displays the expected doughnut shape observed in previous air-plasma experiments and attributed to a Bessel-Gauss mode of the pulse.\cite{Klarskov2013} Here, the observed asymmetry is attributed to astigmatism in the imaging system. \cite{Rasmussen2023}
\\
\indent To estimate the field strength using the pulse energy, we begin by considering the magnitude of the Poynting vector, which describes the rate at which electromagnetic energy flows through space and is given by $S = \epsilon_0 c |E|^2$, \cite{Griffiths2017} where \( S \) is the energy flux (power per unit area), \( \epsilon_0 \) is the permittivity of free space, \( c \) is the speed of light, and \( E \) is the electric field. The total electromagnetic energy is then expressed as:
\begin{equation}
U = \epsilon_0 c \int_T \int_A |E(x, y, t)|^2 \, dx \, dy \, dt
\end{equation}
For a Gaussian profile, the transverse integral simplifies to:

\begin{equation}
U = \frac{\epsilon_0 c \pi \Gamma_x \Gamma_y}{\ln(16)} \int_T |E(t)|^2 \, dt
\end{equation}
Assuming that the electric field can be written as \( E(t) = E_0 f(t) \), where $f(t)$ represents a normalized time-domain trace of the electric field, the total energy can be expressed as:

\begin{equation}
U = \frac{\epsilon_0 c \pi \Gamma_x \Gamma_y}{\ln(16)} E_0^2 \int_T |f(t)|^2 \, dt
\end{equation}

Solving for \( E_0 \), the peak electric field, we get:

\begin{equation}\label{eq:field}
E_0 = \sqrt{\frac{\ln(16)}{\epsilon_0 c \pi \Gamma_x \Gamma_y} \frac{U}{\int_T |f(t)|^2 \, dt}}
\end{equation}

\begin{figure*}[ht]
\centering
\includegraphics[width=\textwidth]{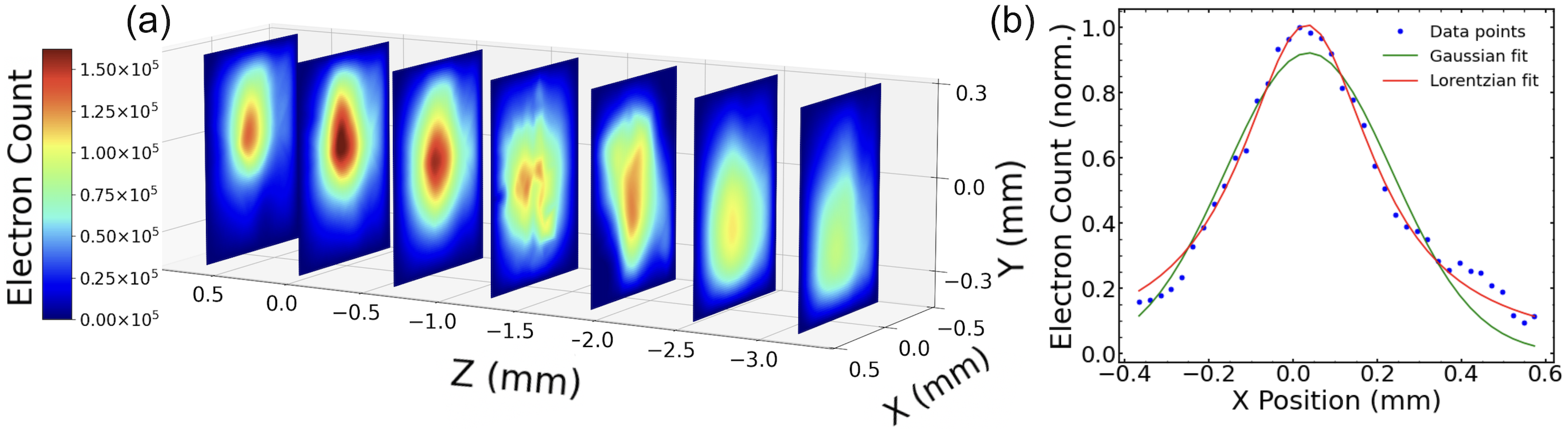} 
\caption{(a) Interpolated electron emission cross sections for different values in Z. (b) Emission line-segment of beam waist as a function of X position with Lorentzian and Gaussian fits superimposed.}
\label{fig:Z_cuts}
\end{figure*}

Using a time-domain trace obtained from air biased coherent detection (ABCD), the envelope of the field is obtained through a Hilbert transform, yielding \( \int_T f^2(t) \, dt = 0.13 \pm 0.01~\text{ps} \). ABCD is preferred for this measurement as it avoids material-dependent limitations of electro-optic crystals, such as frequency-dependent phase mismatch, dispersion, and multiple reflections within the crystal, which can lead to temporal smearing and distortions in EOS \cite{Jepsen2011}. The uncertainty in the time integral of the pulse envelope, \( \int_T f^2(t) \, dt \), was determined by comparing the measured temporal pulse duration to that of a transform-limited Gaussian pulse, which gives a bandwidth-limited duration of \( 0.12~\text{ps} \). We measured the pulse energy as \( U = 186~\text{nJ} \) using a calibrated power meter, which introduces a \( \pm 4\% \) uncertainty, corresponding to \( \pm 7~\text{nJ} \). Using these values, the peak electric field was calculated to be \(\approx 600~\text{kV/cm} \), with lower bound and upper bounds of  \( 400\) and \( 900~\text{kV/cm} \), accounting for uncertainty. These fields, when applied to our nanotip, would be expected to yield electron bunches of approximately $10^7$~e$^-$/pulse with peak energies exceeding  6~keV, motivating the experiment presented here.

Following these initial measurements, unexpected changes were observed in the relative timing of the probe and THz pulses during the optimization of the EO signal and crystal scanning. This suggests the presence of several pulses arriving staggered in time, likely due to different emission times in the plasma or different path lengths traveled by each point-emitter in the plasma filament. This timing behavior has been fully described in a recent paper.\cite{paulino2024spatiotemporal} Here, further evidence of this effect is described and its impact on our observed electron emission. Most importantly to our electron emission experiments, this observation indicates that the field strength provided using the time-averaged Poynting vector above may be significantly overestimated, since the power meter cannot distinguish between a single large pulse or several smaller ones being focused to the optical axis at different points in time and space. As electron emission is an excellent probe of the quasi-static peak THz field, our data will serve as another method of field strength characterization with which to compare these estimates.

In our electron emission experiments, the nanotip position was scanned along the z-axis, revealing an extended, filament like focus, as shown in Fig. \ref{fig:Z_cuts} (a). This observation follows our expectation of a Bessel-Gauss beam described elsewhere.\cite{Klarskov2013} In this figure the data is interpolated, with a total of 243 points for an average of 35 grid values per cross-section. During this scan a maximum of 1.5x$10^5$ electrons per pulse were measured via the replenishment current at the base, though at other times during the experiment as many as 2x$10^5$ were observed. In Fig. \ref{fig:Z_cuts} (b), an electron emission line segment across X is shown, with Gaussian and Lorentzian fits superimposed. The data is better described by a Lorentzian distribution, consistent with previous observations of air-plasma beam waists.\cite{Klarskov2013}

\begin{figure}[ht]
    \centering
    {\includegraphics[width=0.99\linewidth]{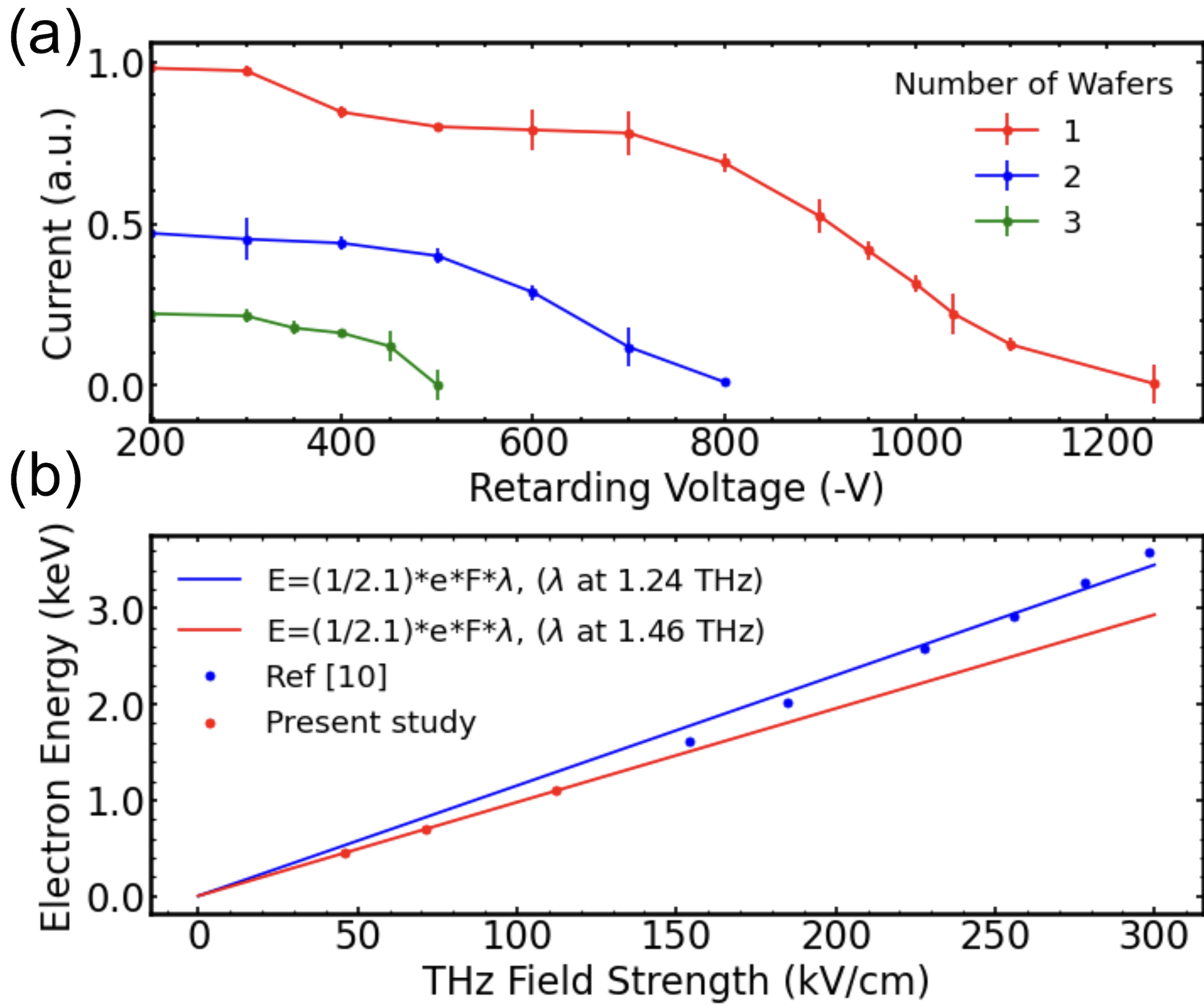}}
    \caption{
    (a) Electron emission current detected at the Faraday cup vs retarding potential with 1, 2, 3 silicon wafers in the THz beam path. (b) Electron peak energy vs incident field strength, with data from the present work in red and data taken from the work of Matte \textit{et al.} \cite{Matte2021} in blue, with corresponding field strengths for peak energies.} 
    \label{fig:energies}
\end{figure}
The electron energy spectrum was obtained by applying retarding voltages to the Faraday cup. Measurements were taken with and without silicon wafers in the collimated THz beam path to modulate the incident field. As shown in Fig. \ref{fig:energies}(a), the maximum electron energies detected were 1100~eV, 700~eV, and 450~eV for 1, 2, and 3 silicon wafers, respectively, far below the  $\approx$ 6~keV expected from a 600~kV/cm source. These measurements were obtained by iteratively querying our electrometer, with each current value representing an integrative measurement of charge over time. The current averages and their standard deviation were calculated across five iterations, resulting in the error bars shown.

Upon tunneling, electrons undergo ballistic acceleration and gain energy proportional to $F_{loc}$, diminishing to $F_0$ over a distance comparable to $R$.\cite{LiJones2016, Miller1996} Consequently, the energy gained by the electrons ($E$) is equivalent to the work done by the local electric field, $ E \sim \frac{1}{2} \gamma e F_0 R$, where \(\gamma = F_{loc}/F_0\) is the field enhancement factor, and \(e\) is the elementary charge of the electron.  This relationship, however, 
implies a large dependence on $R$ for the final electron energies, something which Li \textit{et al.} found to disagree with experimental observations.\cite{LiJones2016} To explain these findings, Li \textit{et al.} used the fact that $\gamma$, for a hemisphere-capped cylinder of length $L$ and radius $R$ in a static field, is proportional to $\sim \frac{L}{R}$. \cite{Podenok2006} This value can be further simplified to remove the length dependence, $L$, as retardation effects will limit the effective length of the emitter to $\lambda/2$.\cite{Kang2009LocalCapacitor} Plugging this into the earlier expression gives: 
\begin{equation}
    \Delta E \sim \frac{1}{4} e F_0 \lambda
    \label{eq:two}
\end{equation}
Giving us direct dependence of incident light field and wavelength on the electron energy, independent of $R$. Using our electron energies and Eq.~\ref{eq:two}, the THz field strength may once again be estimated. As this is a proportionality relationship, it can be compared with existing electron energy vs field strength data to obtain an empirical version. Matte \textit{et al.} performed a similar energy scan, using THz generated by OR in LiNbO$_3$, and determined field strengths via EO sampling and as well as using the Poynting method, displaying excellent agreement. While we also attempted to calibrate the field strength of the plasma emitted THz pulses using EO sampling, the results of this calibration were always much too low to be considered realistic, again perhaps due to the spatio-temporal profile of the THz pulse mentioned above and the much higher bandwidth of the plasma THz pulses. In Fig. \ref{fig:energies} (b), it can be seen that for the data of Matte \textit{et al.}, equation \ref{eq:two} is better approximated by replacing the factor of $1/4$ with a factor of $1/2.1$. Using this relationship, we then estimate the plasma generated THz peak field strengths are 112.4, 71.5, 46.0 ~kV/cm, for 1, 2 and 3 wafers respectively. Accounting for the 70\% transmission of each silicon wafer, the maximum field produced by air-plasma is approximately 160~kV/cm, significantly lower than the previous estimate based on THz power of 600~kV/cm. Given the magnitude of this discrepancy, it is unlikely to be caused by the aforementioned spatio-temporal profile of the THz pulse, suggesting an overestimation of field strength using the Poynting method. 

\begin{figure}[ht]
    \centering
    {\includegraphics[width=0.99\linewidth]{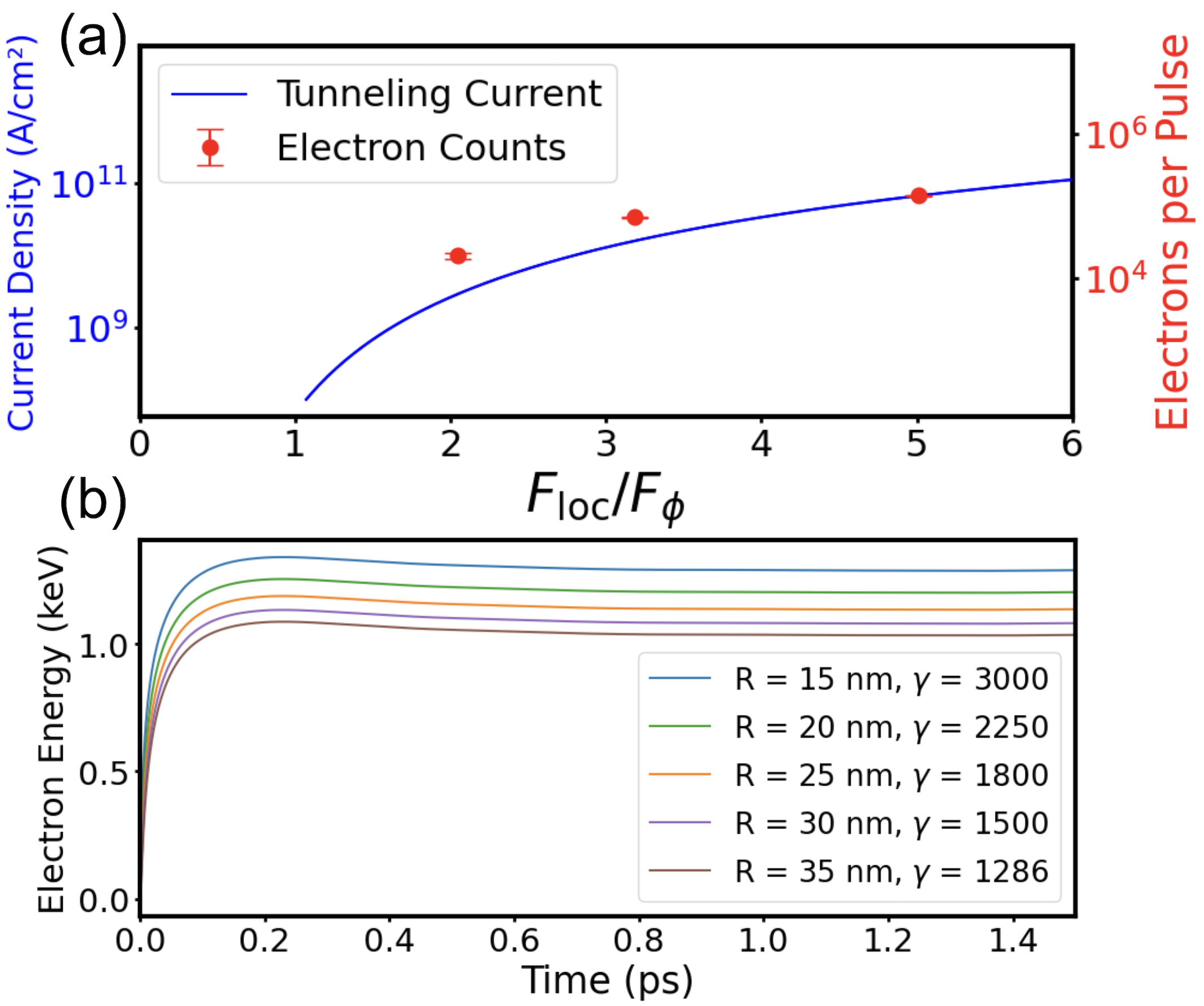}}
    \caption{
    (a) Emitted electrons per pulse and tunneling current as predicted by FN tunneling theory, plotted against the ratio of local field to  the field required to narrow the potential barrier to zero $(F_{loc}/F_{\phi})$. (b) Simulated electron energy as a function of time for various tip radii, with corresponding field-enhancement values ($\gamma$).}
    \label{fig:fowler}
\end{figure}

Fig.~\ref{fig:fowler} (a) shows the measured bunch charges for the THz field strengths investigated, and a comparison to the expectation of FN theory. Assuming an emission duration of 0.4~ps and an emission area of 10~nm$^2$, a field enhancement factor $\gamma$ of 6900 was estimated for a local field of 70~GV/m. This value is significantly higher than similar experiments which reported $\gamma$ values between 800 and 3800.  \cite{LiJones2016,Matte2021, Vella2021}. Potential explanations include variations in the work function due to surface contamination, which might alter the local field and geometry of the tip; deviations in emission area due to changing tip morphology; the presence of multiple emission sites on the tip surface; and space-charge effects, which could have influenced our observed electron energy distributions by causing additional broadening. However, these effects alone are unlikely to fully account for the observed discrepancies.

Another possible explanation is the presence of multiple pulses arriving staggered in time. To further understand this, the ballistic acceleration of electrons in the enhanced illuminating field was simulated. Assuming the near-field fall-off given by $\gamma(x) = \gamma/(1+2x/R)$ for a hyperboloid shaped tip,\cite{LiJones2016, Miller1996} the dynamics of our highest energy electrons are modeled, emitted at the peak electric field, thus undergoing the maximum possible acceleration. In this purely 1D simulation, the electron's position and velocity are iteratively computed and updated at each time step. The code calculates the electric field at the current position, based on spatial decay and the passing pulse, to determine the force acting on the electron. The simulated temporal evolution of the electron energy is shown in Fig. ~\ref{fig:fowler} ~(b). The field enhancement factor ($\gamma$) importantly scales as $\sim \lambda/R$.\cite{Kim2008} As our tip radius is not precisely known, a range of tip radii are given, with a range of enhancement factors then found. Our simulations show that, for tips between 15-35~nm in radii and assuming a incident field of 110~kV/cm estimated above, a field enhancement between $1300$ to $3000$ could have produced our maximum detected electron energies of $\sim 1.1$~keV. This range is closer to our expectation based on similar experiments. We note that although the $\gamma$ estimate based on electron energy is directly related to the highest incident field on the tip and thus remains unaffected by the presence of multiple staggered pulses, the same cannot be said for our electron counts. In the previous FN-based estimation of $\gamma$, it was assumed that the tip is illuminated by a single pulse with a peak field of 110~kV/cm every 1 ms, given by repetition rate of our laser. However, if this assumption is incorrect—such as in the case where multiple pulses with weaker fields arrive more frequently, our calculated $\gamma$ would be significantly reduced, bringing it in line with the simulated values and expectation.

In conclusion, sub-cycle cold field electron emission from a tungsten nanotip using a two-color laser plasma was studied, achieving electron energies of 1.1~keV and bunch charges up to \(2 \times 10^5\) electrons per pulse. 
While THz peak field strengths up to 600~kV/cm were estimated using the Poynting flux method, observations were consistent with a much lower field of 160~kV/cm based on the relationship between incident field and electron energy. This discrepancy was attributed to spatio-temporal focusing effects arising from collimation in the long plasma filaments using standard off-axis parabolic mirrors, producing a flying THz focus.\cite{paulino2024spatiotemporal} 

We note that preliminary results using a plastic axicon as a collimating optic showed great promise, with a more than fourfold increase in peak THz field compared to the standard off-axis parabolic mirror, suggesting THz fields of over 600~kV/cm. By adopting this collimation scheme for future experiments, the current electron emission record of \(10^6\)~e$^-$/pulse is expected to be surpassed, promising bunch charges up to $10^7$~e$^-$/pulse and peak energies exceeding  6~keV. These advancements position plasma-driven THz sources as a promising option for ultrafast electron emission and possible THz-driven ultrafast electron microscopy using  high-brightness, coherent electron beams from metal nanotips.
\\
\\
The authors wish to acknowledge financial support from the Natural Sciences and Engineering Research Council of Canada (NSERC), Fonds de Recherche du Québec-Nature et Technologies (FRQNT) and the Canadian Foundation for Innovation (CFI).
\nocite{*}
\section*{AUTHOR DECLARATIONS}
\subsection*{Conflict of Interest}
\vspace{-3mm}
The authors have no conflicts to disclose.
\vspace{-3mm}
\subsection*{Author Contributions}
\vspace{-3mm}
\textbf{Benjamin Colmey}: Conceptualization (equal); Formal analysis (equal); Investigation (lead); Validation (equal); Writing – original draft (lead); Writing – review and editing (lead). \textbf{Rodrigo T. Paulino}: Conceptualization (equal); Formal analysis (equal); Investigation (equal); Validation (equal); Writing – review and editing (equal). \textbf{Gaspard Beaufort}: Investigation (supporting). \textbf{David G. Cooke}: Conceptualization (equal); Formal analysis (equal); Funding acquisition (lead); Project administration (lead); Validation (lead); Supervision (lead); Writing – review and editing (equal).
\vspace{-3mm}
\section*{Data Availability Statement}
\vspace{-1mm}
The data that support the findings of
this study are available from the
corresponding author upon reasonable
request.
\vspace{-3mm}
\section*{References}
\vspace{-5mm}

\bibliography{aipsamp}

\end{document}